\documentclass[twocolumn,prb]{revtex4}
\usepackage{graphicx,amsmath}

\begin{document}

\title{Nonlinear screening and percolative transition in a two-dimensional
electron liquid}

\author{Michael M. Fogler}

\affiliation{Department of Physics, University of
California San Diego, La Jolla, California 92093}

\date{\today}

\begin{abstract}

A variational method is proposed for calculating the percolation
threshold, the real-space structure, and the ground-state energy of a
disordered two-dimensional electron liquid. Its high accuracy is
verified against exact asymptotics and prior numerical results. The
inverse thermodynamical density of states is shown to have a strongly
asymmetric minimum at a density that is approximately the triple of the
percolation threshold. This implies that the experimentally observed
metal-insulator transition takes place well before the percolation point
is reached.

\end{abstract}
\maketitle

\begin{figure}
\centerline{
\includegraphics[width=2.1in]{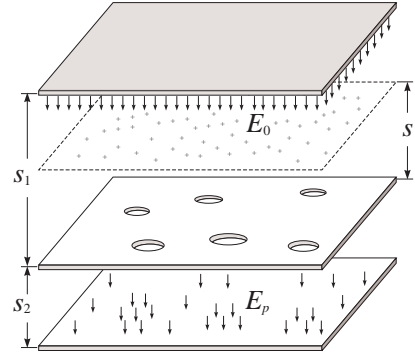}
}
\vspace{0.1in}
%\setlength{\columnwidth}{3.2in}
%\centerline{
\caption{
The geometry of the theoretical model. The 2D layer of interest is
sandwiched between the top and the bottom metallic gates. The dopants
reside in the plane with the dashed border. The depletion regions (shown
as holes in the probed layer) enhance the penetrating electric field
$E_p$.
\label{Fig_bilayer}
}
%}
\end{figure}

The discovery that a two-dimensional (2D) electron liquid can be a metal
at moderate electron density $n_e$ and an insulator at small $n_e$ was a
major surprise that questioned our fundamental understanding of the role
of disorder in such systems. Today, almost a decade later, it remains a
subject of an intense debate.~\cite{Abrahams_01} One important reason
why the conventional theory fails could be its flawed basic premise of
the ``good'' metal, i.e., a uniform electron liquid slightly perturbed
by impurities and defects. Indeed, modern nanoscale imaging
techniques~\cite{Eytan_98,Finkelstein_00,Ilani_00,Morgenstern_00}
unambiguously showed that low-density 2D electron systems are strongly
inhomogeneous, ``bad'' metals, where effects of disorder are
nonperturbatively strong. In particular, depletion regions (DR), i.e.,
regions where $n(\textbf{r})$ is effectively zero, exist. They appear
when $n_e$ is too small to adequately compensate fluctuating charge
density of randomly positioned impurities. As $n_e$ is reduced in the
experiment, e.g., in order to approach the vicinity of the
metal-insulator transition (MIT), the DRs are expected to grow in size
and concentration and eventually merge below some percolation threshold
$n_e = n_p$. An important and controversial issue is whether or not this
percolation transition plays any role in the observed
MIT.~\cite{Abrahams_01} To resolve it one needs to have a theory that is
able to calculate $n_p$ and that can describe the inhomogeneous
structure of the 2D metal at $n_e \sim n_p$. A great progress in this
direction has been achieved by Efros, Pikus, and
Burnett,~\cite{Efros_93} whose paper is intellectually tied to earlier
work on nonlinear screening by Efros and
Shklovskii.~\cite{Efros_Shklovskii_book} Still, analytical results
remained scanty and numerical simulations~\cite{Efros_93,Shi_02} were
the only known way to quantitatively study the $n_e \sim n_p$ regime.
These simulations are very time consuming and redoing them in order to
get any information beyond what is published~\cite{Efros_93,Shi_02} or
to study novel experimental setups seems impracticable. Below I will
show that a variational approach to the problem can be a viable
alternative. Comparing it with the available numerical results for a
typical model of the experimental geometry (Fig.~\ref{Fig_bilayer}), I
establish that it correctly predicts the value of $n_p$ and accurately
reproduces the energetics of the ground state, in particular, the
density dependence of the electrochemical potential $\mu$ and of the
inverse thermodynamical density of states (ITDOS), $\chi^{-1} = d \mu /
d n_e$, over a broad range of $n_e$. The most striking feature of the
resultant function $\chi^{-1}(n_e)$ is a strongly asymmetric minimum,
which is observed in real
experiments.~\cite{Eisenstein_94,Dultz_00,Dultz_xxx,Allison_02} I will
elaborate on the origin of this feature and show that it occurs at $n_e
\approx 3 n_p$ largely independently of the parameters of the system.
Recently, this minimum attracted much interest when Dultz \textit{et
al.\/}~\cite{Dultz_00} reported that in some samples it virtually
coincides with the apparent MIT. The proposed theory indicates that, at
least for these samples, any connection between the MIT and the
percolation of the DRs can be ruled out. Thus, the explanation of the
MIT lies elsewhere.

%
% FIG. 1
%

The Hamiltonian of the model is adopted from Efros, Pikus, and Burnett
(EPB)~\cite{Efros_93} (see also Fig.~\ref{Fig_bilayer}),
\begin{equation}
H =  \int d^2 r \left\{\frac12 [n(\textbf{r}) - n_e] \Phi(\textbf{r})
+ H_0(n)\right\}
\label{H}
\end{equation}
where $\Phi(\textbf{r})$ is the electrostatic potential,
\begin{equation}
\Phi = \int d^2 r^\prime \frac{e^2}{\kappa} \left[
         \frac{n(\textbf{r}^\prime) - n_e}{|\textbf{r}^\prime - \textbf{r}|}
      -  \frac{n_d(\textbf{r}^\prime) - n_i}
              {\sqrt{(\textbf{r}^\prime - \textbf{r})^2 + s^2}}\right]
\label{Phi}
\end{equation}
$\kappa$ is the dielectric constant, and $H_0(n)$ is the energy density
of the uniform liquid of density $n$,
\begin{equation}
               H_0(n) = - (e^2  / \kappa) n^{3/2} h_0(n).
\label{H_0}
\end{equation}
At low densities, $h_0(n) \sim 1$ is a slow function~\cite{Tanatar_89} of
$n$. The negative sign in Eq.~(\ref{H_0})
reflects the prevalence of the exchange-correlation energy over the
kinetic one in this regime.
The potential created by the random concentration of dopants $n_d({\bf
r})$ can be expressed in terms of the effective \textit{in-plane\/}
background charge $\tilde{\sigma}(\textbf{q}) = \tilde{n}_d(\textbf{q})
\exp(-q s)$ (the tildes denote the Fourier transforms). With these
definition, $\Phi(\textbf{r})$ coincides with the potential created by the
charge density $\sigma(\textbf{r}) = n(\textbf{r}) - n_e - \sigma(\textbf{r})$. I will
assume that the dopants have the average concentration $\langle n_d
\rangle \equiv n_i \gg s^{-2}$ and are uncorrelated in space. In this
case the one- and two-point distribution functions of $\sigma(\textbf{r})$ have
the Gaussian form,
\begin{eqnarray}
&P_1(\sigma) = (2 \pi K_0)^{-1/2} \exp (-\sigma^2 / 2 K_0),\quad&
\label{P_1}\\
&\displaystyle P_2 = \frac{1}{2 \pi \sqrt{K_0^2 - K_{r^\prime}^2}}
\exp\left[\frac{2 K_{r^\prime} \sigma \sigma^\prime - K_0 (\sigma^2 + \sigma^{\prime\,2})}
               {2 (K_0^2 - K_{r^\prime}^2)}
    \right]\quad&
\label{P_2}
\end{eqnarray}
where $\sigma^\prime = \sigma(\textbf{r} + \textbf{r}^\prime)$ and $K_r$ is given by
\begin{equation}
   K_r \equiv \langle \sigma(r) \sigma(0) \rangle  = {n_i s} /
   \pi {(r^2 + 4 s^2)^{3/2}}.
\label{K}
\end{equation}
The characteristic scales in the problem are as follows.~\cite{Efros_93}
The typical amplitude of fluctuations in $\sigma$ is $\sqrt{n_i} / s$, see
Eqs.~(\ref{P_1}) and (\ref{K}). Their characteristic spatial scale is
the spacer width $s$ [cf.~Eqs.~(\ref{P_1}--\ref{K}) and
Fig.~\ref{Fig_bilayer}]. In the cases studied below, $n_e \gtrsim n_p$
and $n_e \gg n_p$, $s$ exceeds the average interelectron separation $a_0
= n_e^{-1/2}$. As usual in Coulomb problems, the energy is dominated by
the longest scales, in this case $s$. The fluctuations $H_0(n) -
H_0(n_e)$ of the the local energy density come from the interactions on
the much shorter scale of $a_0 \ll s$ and can be treated as a
perturbation.~\cite{Efros_93} For the purpose of calculating the
ground-state density profile $n(\textbf{r})$ I neglect $H_0$. Once such a
ground state is known, I correct the total energy $H$ by adding to it
$H_0$ averaged over $n(\textbf{r})$.

To find $n(\textbf{r})$ one needs to solve the electrostatic problem
with the following dual boundary conditions: if $n(\textbf{r}) > 0$,
then $\Phi(\textbf{r}) = \mu = {\rm const}$; otherwise, if $n = 0$ (DR),
then~\cite{Efros_Shklovskii_book,Efros_93,Fogler_comp_cb} $\Phi > \mu$.
I start with the analysis of the large-$n_e$ case, which clarifies why
$\chi^{-1}(n_e)$ dependence is nonmonotonic and which provides a formula
for the the density $n_m$ where $\chi^{-1}$ has the minimum.

In the limit $n_e \gg \sqrt{n_i} / s$, the asymptotically exact
treatment is possible because the DRs appear only in rare places where
$\sigma(\textbf{r})$ dips below $-n_e$. (In practice, this limit is
realized when $n_e > K_0^{1/2} \approx 0.2 \sqrt{n_i} / s$). The
corresponding electrostatic problem is analogous to that of the metallic
sheet perforated by small holes, see Fig.~\ref{Fig_bilayer}. The most
elegant way to derive $\chi^{-1}$ in this regime is to calculate the
fraction $E_p / E_0$ of the electric field that reaches the bottom layer
in the geometry of Fig.~\ref{Fig_bilayer}. Indeed, $E_p / E_0$ is
nonzero only if the probe layer is not a perfect metal, $\chi^{-1} \neq
0$. In the simplest case, where distances $s_1$ and $s_2$ are large, the
following formula holds: $\chi^{-1} = 4 \pi (e^2 s_2 / \kappa)(d E_p / d
E_0)$ (cf.~Ref.~\onlinecite{Eisenstein_94}). This is essentially the
formula used to deduce $\chi^{-1}$ in the
experiment~\cite{Eisenstein_94,Dultz_00,Dultz_xxx,Allison_02}). It is
immediately obvious that holes in the metallic sheet enhance the
penetrated field $E_p$. For example, the field leaking through a round
hole of radius $a$ is the field of a dipole~\cite{Jackson_book}
$\textbf{p} = (a^3 / 3 \pi) \textbf{E}_0$. If there is a finite but
small concentration $N_h$ of such holes, their fields are additive,
leading to $\chi^{-1} = (8 \pi e^2 / 3 \kappa) N_h a^3$. From here the
exact large-$n_e$ asymptotics of $\chi^{-1}$ and $\mu$ can be obtained
by substituting the proper $N_h$ and averaging over the distribution of
$a$. This can be done by noting that these holes appear around the
minima of $\sigma(\textbf{r})$ whose statistics is fixed by
Eqs.~(\ref{P_1}--\ref{K}). It is easy to see that the most probable
holes are nearly perfect circles with
radii~\cite{Efros_93} $a \sim s \sqrt{K_0} / n_e$.
The charge distribution around a single hole at distances $r > a$ is
given by the formula
\begin{equation}
n(r) = \frac{\sigma_{x x} a^2}{\pi}
\left[ \sqrt{\frac{r^2}{a^2} - 1} - \left(\frac{r^2}{a^2} +
\frac{2}{3}\right) \arccos \frac{a}{r}\right],\:
\label{sigma_hole_out}
\end{equation}
where the hole is assumed to be centered at $r = 0$ and $a^2 = -3 [n_e +
\sigma(0)] / \sigma_{x x}$. Equation~(\ref{sigma_hole_out}) can be
obtained, e.g., by generalizing the textbook
solution~\cite{Jackson_book} for the hole in the metallic
sheet~\cite{Comment_on_hole} and is also the limiting form of Eq.~(11) in
Ref.~\onlinecite{Burnett_93}. Previously, Eq.~(\ref{sigma_hole_out}) was
used for study of quantum dots in Ref.~\onlinecite{Fogler_94}.

In the current problem the main factor that determines the net
contribution $\chi^{-1}_{\rm DR}$ of the depletion holes to $\chi^{-1}$
is their exponentially small concentration, proportional to $P_1(-n_e)
\propto \exp(-n_e^2 / 2 K_0)$. The final result,
\begin{equation}
 \chi^{-1}_{\rm DR} \simeq ({3 \sqrt{2}}/{8 \pi})
 ({e^2 n_i}/{\kappa s n_e^2})
 \exp(-{4 \pi s^2 n_e^2}/{n_i}),
\label{chi_asymp}
\end{equation}
agrees with that of EPB but has no numerical coefficients left
undetermined. To finalize the calculation, one needs to augment
$\chi^{-1}_{\rm DR}$ by the local term $\chi_0^{-1} = d^2 \langle H_0(n)
\rangle / d n_e^2$. In the present case, fluctuations around the average
density are small. Hence, $\langle H_0(n) \rangle \simeq H_0(n_e)$ and
\begin{equation}
 \chi^{-1}_0(n_e) \simeq -(e^2 / \kappa) h_1(n_e) / \sqrt{n_e},\quad
  n_e \gg \sqrt{n_i} / s,
\label{chi_0_large_n}
\end{equation}
where $h_1(n_e) = (3 / 4) h_0(n_e) + 3 h_0^\prime n_e + h_0^{\prime\prime}
n_e^2 \sim 1$.

\begin{figure}
\centerline{
\includegraphics[width=2.5in,bb=145 150 450 670]{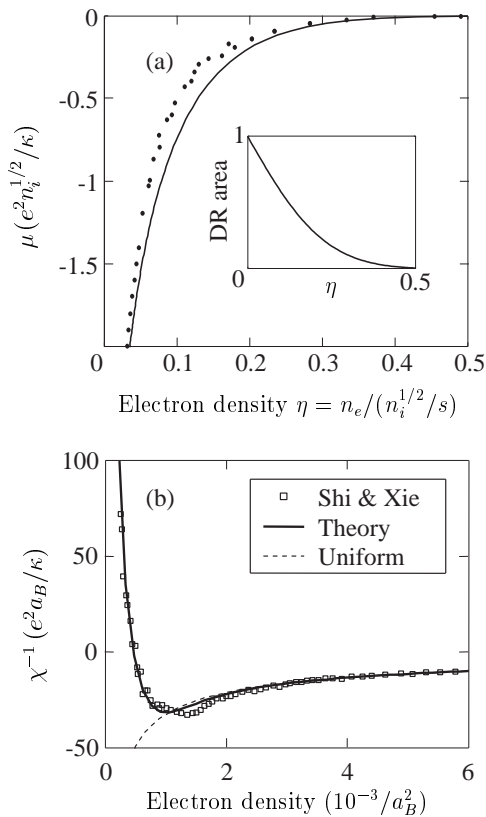}
}
\vspace{0.1in}
\setlength{\columnwidth}{3.2in}
%\centerline{
\caption{
(a) Electrochemical potential \textit{vs.\/} electron density for the
electrostatic problem, $H_0 \equiv 0$, according to the present theory
(solid line) and EPB's numerical simulations (dots). Inset: Fraction of
the depleted area \textit{vs.\/} density. (b) $\chi^{-1}$ {\it
vs.\/} density according to the numerical simulations of Shi and
Xie~\cite{Shi_02} (squares), the present theory (solid line), and for
the uniform electron liquid (dashed line).
\label{Fig_comparison}
}
%}
\end{figure}

Formula~(\ref{chi_asymp}) implies a sharp upturn of the ITDOS as $n_e
\to 0$ caused by the exponential growth of the DRs. At the boundary of
its validity, $n_e \sim \sqrt{n_i} / s$, Eq.~(\ref{chi_asymp}) gives
$\chi^{-1}_{\rm DR} \sim e^2 s / \kappa$, which is large and
\textit{positive\/}. On the other hand, the local term $\chi^{-1}_0$
[Eq.~(\ref{chi_0_large_n})] shows a weak dependence on $n_e$, remaining
small and \textit{negative\/}. Combined, they produce a strongly
asymmetric minimum in $\chi^{-1}(n_e) = \chi^{-1}_{\rm DR} +
\chi^{-1}_0$ at the density
\begin{equation}
 n_m = \frac{1}{4 \sqrt{\pi}} \frac{\sqrt{n_i}}{s}
\ln^{1/2} \left(\frac{4096}{\pi h_1^4} n_i s^2\right).
\label{n_m}
\end{equation}
The log-factor in Eq.~(\ref{n_m}) is rather insensitive to $n_i s^2$
and $h_1$. For $n_i s^2 = h_1 = 1$, one gets $n_m \approx
0.38 {\sqrt{n_i}}/\,{s}$. In comparison, (see Ref.~\onlinecite{Efros_93}
and below) the percolation threshold is
\begin{equation}
                    n_p \approx 0.12 {\sqrt{n_i}}/\,{s},
\label{n_p}
\end{equation}
so that $n_m \approx 3 n_p$. In accord with EPB's heuristic argument, at
such density a very small area fraction is depleted (see inset in
Fig.~\ref{Fig_comparison}a), and so the use of the asymptotic
formula~(\ref{chi_asymp}) is justified.

%
% FIG. 2
%

Let us now proceed to the case $n_e \sim n_p$. Again, I start with the
electrostatic problem ($H_0 \equiv 0$). One expects DRs to be abundant
and irregularly shaped. For a given $n_e$, the ground-state $n(\textbf{r})$
is some nonlocal functional of $\sigma$ and there is no hope to find it exactly.
What I wish to report here is that a variational solution sought within
the class of purely local functionals, $n(\textbf{r}) = n[\sigma(\textbf{r})]$
remarkably accurately reproduces the $\mu(n_e)$ and $\chi^{-1}(n_e)$
dependencies found in numerical work.~\cite{Efros_93,Shi_02} More
interestingly, it predicts the correct value of $n_p$ [Eq.~(\ref{n_p})].

The system of equation that defines such a variational state
$n(\sigma)$ follows from the fact that for a Gaussian random
function the averages over the total area $L^2$ of the system and over
the distribution function are the same. This yields
[cf.~Eqs.~(\ref{H}--\ref{P_2})]
\begin{eqnarray}
&\displaystyle \frac{H_v}{L^2} = \frac12 \int \int d \sigma d \sigma^\prime
\rho(\sigma) G(\sigma, \sigma^\prime) \rho(\sigma^\prime),\quad\quad&
\label{H_v}\\
&\displaystyle G(\sigma, \sigma^\prime) =  \int\limits d^2 r^\prime V(r^\prime)
                 [P_2(\sigma, \sigma^\prime) - P_1(\sigma) P_1(\sigma^\prime)],\quad\quad&
\label{G}
\end{eqnarray}
where $\rho(\sigma) = n(\sigma) - n_e - \sigma$ and $V(r^\prime) = e^2 / \kappa
r^\prime$. The energy $H_v$
needs to be minimized with respect to all functions $n(\sigma)$
that obey the constraints $n(\sigma) \geq 0$ and
\begin{equation}
 \int d \sigma P_1(\sigma) n(\sigma) = n_e.
\label{f_condition}
\end{equation}
The latter ensures that the average density is equal to $n_e$.
Introducing the Lagrange multiplier $\mu_v$ (variational estimate of the
electrochemical potential), one obtains that $H_v$ is minimized
if, for all $\sigma > f$, $\rho(\sigma^\prime)$ satisfies
\begin{equation}
\int\limits d \sigma^\prime
 G(\sigma, \sigma^\prime) \rho(\sigma^\prime) = \mu_v(n_e) P_1(\sigma)
\label{var_equation}
\end{equation}
Here $f$ is such that $n(f) = 0$. [I found that $n(\sigma)$ is always a
monotonically increasing function, so that $n > 0$ corresponds to
$\sigma > f$]. The kernel $G(\sigma, \sigma^\prime)$ [Eq.~(\ref{G})] is
log-divergent, $G \propto -\ln|\sigma - \sigma^\prime|$ at $\sigma \to
\sigma^\prime$ and decays exponentially at large $\sigma$ and
$\sigma^\prime$. There is a certain analogy between
Eq.~(\ref{var_equation}) and the integral equations of 1D
electrostatics, which also have log-divergent
kernels.~\cite{Jackson_book,Fogler_94} This analogy entails that
$n(\sigma) \sim \sqrt{\sigma - f}$ at $\sigma$ close to $f$. Note that
$\sigma(\textbf{r}) - f$ is proportional to the distance in the real
space between the given point $\textbf{r}$ and the boundary of the
nearby DR. Thus, the variational principle renders correctly the
square-root singularity in $n(\textbf{r})$ at the edge of the DRs [cf.,
e.g.,~Eq~(\ref{sigma_hole_out})]. I was not able to establish the
analytical form of the solution beyond this property and resorted to
finding $n(\sigma)$ numerically. To do so the integral in
Eq.~(\ref{var_equation}) was converted into a discrete sum over 101
points on the interval $|\sigma| < 1.5 \sqrt{n_i} / s$ the resultant
system of 101 linear equations was solved on the computer. The solution
can be approximated by a simple analytical \textit{ansatz\/}
\begin{equation}
 n_a(\sigma) = [(n_e + \sigma)^2 - (n_e + f)^2]^{1/2} \theta(\sigma - f),
\label{n_var}
\end{equation}
where $\theta(z)$ is the step-function. For example, the energy $H_v$
is nearly the same whether it is calculated using $n_a$ or using the
actual solution of Eq.~(\ref{var_equation}). So, in principle,
Eq.~(\ref{n_var}) obviates the need to solve Eq.~(\ref{var_equation}).
The only equation that needs to be solved is Eq.~(\ref{f_condition}) for
$f$.

At this point one can compare the predictions of the variational
method for $\mu_v(n_e)$ with EPB's numerical results that were also
obtained for the $H_0 = 0$ problem. As Fig.~\ref{Fig_comparison}a
illustrates, they are in a good agreement.~\cite{Comment_on_ansatz}

To test the theory further I compare it next with the numerical data of
Shi and Xie.~\cite{Shi_02} To this end, one needs to calculate
$\chi^{-1}$ including the effect of a finite $H_0$. The first step is to
take the derivative $\chi^{-1}_{\rm DR} = d \mu_v / d n_e$, which is
easily done numerically. An accurate fit to the result is provided by
the interpolation formula
\begin{equation}
\chi^{-1}_{\rm DR} \approx \frac{e^2 s}{\kappa}
\frac{3 \sqrt{2}}{8 \pi \eta}
\frac{0.30 + \eta}{0.036 + 0.12 \eta + \eta^2}
\exp(-4 \pi \eta^2),
\label{chi_DR_interp}
\end{equation}
where $\eta \equiv n_e s / \sqrt{n_i}$. This particular form is devised
to match Eq.~(\ref{chi_asymp}) at large $n_e$ and is consistent up to
log-corrections with the behavior
expected~\cite{Efros_Shklovskii_book,Gergel_78} at very small $n_e$. The
next step is to evaluate the quantity 
\begin{equation}
\chi^{-1}_0 = \frac{d^2}{d n_e^2} \langle H_0 \rangle
= -\frac{d^2}{d n_e^2} \int_f^\infty d \sigma P_1(\sigma) H_0[n(\sigma)],
\label{chi_0}
\end{equation}
which is also easily done on the computer. The total ITDOS,
$\chi^{-1}(n_e) = \chi^{-1}_{\rm DR} + \chi^{-1}_0$, calculated from
Eqs.~(\ref{chi_DR_interp}--\ref{chi_0}) and the theoretical value of
$n_m \sim 0.7 \times 10^{-3} a_B^{-2}$ [per Eq.~(\ref{n_m})] compare
very well with the simulations, see Fig.~\ref{Fig_comparison}b. The
parameters I used are $n_i = 6.25 \times 10^{-4} a^{-2}_B$ and $s = 10
a_B$, where $a_B$ is the effective Bohr radius.~\cite{Comment_on_Shi} In
agreement with the exact results presented above, $\chi_0^{-1} \ll
\chi_{\rm DR}^{-1}$ at $n_e < n_m$, and so the upturn of $\chi^{-1}$ at
low $n_e$ is driven the growth of DRs.

Within the variational method the boundaries of the DRs coincide with
the level lines $\sigma(\textbf{r}) = f$ of the zero-mean random
function $\sigma$. Consequently, the DRs percolate at $f \geq 0$.
Solving Eq.~(\ref{var_equation}) with $f = 0$, one arrives at
Eq.~(\ref{n_p}), which is in excellent agreement with EPB's result $n_p
\approx 0.11 \sqrt{n_i} / s$. One important quantity not reported in the
published numerical works~\cite{Efros_93,Shi_02} is the area fraction of
the DR. Within the variational method, it is equal to ${\rm erfc}(-f /
\sqrt{2 K_0}) / 2$, where ${\rm erfc}(z)$ is the complementary error
function. It is exactly $1 / 2$ at $n_e = n_p$ and increases as $n_e \to
0$ as shown in Fig.~\ref{Fig_comparison}a (inset).

Let us now compare our results with the experimental data. Taking a
rough number $n_i = 3 \times 10^{11} {\rm cm}^{-2}$ and a typical spacer
width $s = 40\,{\rm nm}$ in Eq.~(\ref{n_m}), one gets $n_m \approx 5.2
\times 10^{10} {\rm cm}^{-2}$, in agreement with observed
values.~\cite{Dultz_00} The estimate for the percolation point is $n_p
\approx n_m / 3 \approx 1.7 \times 10^{10} {\rm cm}^{-2}$. Despite some
small uncertainty in the last number (due to the uncertainties in $n_i$
and $h_1$), $n_p$ and $n_m$ differ substantially, and so they can be
easily distinguished experimentaly. Therefore, the observed apparent
MIT~\cite{Dultz_00} at $n = n_m$ has nothing to do with the percolation
of the DRs and moreover with breaking of the electron liquid into
droplets (the latter occurs at $n_e \ll n_p$). From
Fig.~\ref{Fig_comparison}a, one can estimate that the DR occupy mere 6\%
of the total area at $n = n_m$.

One qualitative prediction that follows from Eqs.~(\ref{chi_0_large_n})
and (\ref{chi_DR_interp}) is that the upturn of $\chi^{-1}(n_e)$ should
be sharper in samples with larger $s$, which seems to be the case if the
data of Ref.~\onlinecite{Eisenstein_94} ($s = 14\,{\rm nm}$) are compared with
those of Refs.~\onlinecite{Dultz_00,Dultz_xxx,Allison_02} ($s \leq 4\,{\rm
nm}$). Detailed fits are left for future.

I conclude with mentioning some other theoretical work on the subject.
It was suggested~\cite{Chakravarty_99} that near the MIT, function
$\chi^{-1}(n_e)$ may contain both regular and singular parts. My theory
can be considered the calculation of the former. Its detailed comparison
with experiment may furnish an estimate of the putative singular term.
The effect of disorder on $\chi^{-1}$ was also studied in
Refs.~\onlinecite{Si_98} and~\onlinecite{Asgari_02} but the electron
density inhomogeneity was not accounted for. Finally, a nonmonotonic
$\chi^{-1}(n_e)$ dependence was found in a model~\cite{Orozco_03}
without disorder.

This work is supported by C.~\& W.~Hellman Fund and A.~P.~Sloan
Foundation. I am grateful to B.~I.~Shklovskii for indispensable comments
and insights, and also to H.~W.~Jiang, A.~K.~Savchenko, and J.~Shi for
discussions.

%------------------------------------------------------------------------%


\begin{thebibliography}{99}
\bibitem{Comment_on_hole} Note that $E_0 = 4 \pi \sigma(0) / \kappa +
\text{const}$ and that $p$ can be read off the
asymptotic law $2 \pi \sigma(r) \sim -p / r^3$ valid at $r \gg a$. This
leads to the same formula for $\chi^{-1}$ as in the case of a hole in
a metallic sheet, see more details in Ref.~\onlinecite{Fogler_comp_cb}.
\bibitem{Comment_on_ansatz} This is also true in serveral other geometries,
see Ref.~\onlinecite{Fogler_comp_cb}.
\bibitem{Comment_on_Shi} The data in Ref.~\onlinecite{Shi_02} must be
divided by two to correct a notational error (J.~Shi, private
communication); $n_i$ is four times smaller than in
Ref.~\onlinecite{Shi_02} to account for a peculiar choice $q = e / 2$ of
the impurity charge made there.
\end{thebibliography}
\end{document}